\RequirePackage{fix-cm}
\documentclass[natbib,smallextended]{svjour3}       
\smartqed  

\usepackage{graphicx}

 \journalname{ISSI Book on TDEs}
\usepackage{xspace}

\newcommand{\gammachap}{Gamma-ray Chapter\xspace}
\newcommand{\optchap}{Optical Chapter \citep{vanVelzen20_ISSI}\xspace}
\newcommand{\radiochap}{Radio Chapter \citep{Alexander20}\xspace}

\newcommand{\impostchap}{Imposters Chapter \citep{Zabludoff21_ISSI} \xspace}

\newcommand{\emischap}{Emission Mechanisms Chapter \citep{Roth20_ISSI}\xspace}

\usepackage{ref_macros} 

\usepackage{array}
\newcolumntype{P}[1]{>{\centering\arraybackslash}p{#1}}
\usepackage[flushleft]{threeparttable}
\usepackage{booktabs}

\begin{document}

\title{Reverberation in tidal disruption events: dust echoes, coronal emission lines, multi-wavelength cross-correlations, and QPOs}

\titlerunning{Reverberation in tidal disruption events}        
 
\author{Sjoert van Velzen         
        \and
        Dheeraj R. Pasham 
        \and
        Stefanie Komossa
        \and 
        Lin Yan
        \and
        Erin A. Kara
}

\authorrunning{van Velzen, Pasham, Komossa, Yan, \& Kara} 

\institute{S. van Velzen \at
              Center for Cosmology and Particle Physics, New York University, New\,York, NY\,10003, USA and \\
              Leiden Observatory, Leiden University, Postbus 9513, 2300 RA Leiden, The Netherlands
              \email{sjoert@strw.leidenuniv.nl} \\[-13pt]
        \and
            P.\,R. Pasham  \at
            IT Kavli Institute for Astrophysics and Space Research, Cambridge, MA 02139, USA \\[-13pt]
        \and 
            S. Komossa \at
            Max-Planck-Institut f{\"u}r Radioastronomie, Auf dem H{\"u}gel 69, D-53121 Bonn, Germany \\[-13pt]
        \and
            L. Yan \at
            Caltech Optical Observatories, Cahill Center for Astronomy and Astrophysics, California Institute of Technology, Pasadena, CA 91125, USA \\[-13pt]
        \and
            E. Kara \at
            Kavli Institute for Astrophysics and Space Research, Massachusetts Institute of Technology, 77 Massachusetts Avenue, Cambridge, MA 02139, USA \\[-13pt]
}

\date{}

\maketitle

\begin{abstract}
Stellar tidal disruption events (TDEs) are typically discovered by transient emission due to accretion or shocks of the stellar debris. Yet this luminous flare can be reprocessed by gas or dust that inhabits a galactic nucleus, resulting in multiple reverberation signals. Nuclear dust heated by the TDE will lead to an echo at infrared wavelengths (1-10 $\mu$m) and transient coronal lines in optical spectra of TDEs trace reverberation by gas that orbits the black hole. Both of these signal have been detected, here we review this rapidly developing field. We also review the results that have been extracted from TDEs with high-quality X-ray light curves: quasi periodic oscillations (QPOs), reverberation lags of fluorescence lines, and cross-correlations with emission at other wavelengths. The observational techniques that are covered in this review probe the emission from TDEs over a wide range of scales: from $\sim 1$~light year to the innermost parts of the newly formed accretion disk. They provide insights into important properties of TDEs such as their bolometric output and the geometry of the accretion flow. While reverberation signals are not detected for every TDE, we anticipate they will become more commonplace when the next generation of X-ray and infrared instruments become operational. 

\end{abstract}

\section{Introduction}
\label{intro}
At the time of writing, the field of tidal disruption event (TDE) research finds itself in an interesting position: most of the new events are found in optical surveys, see the \optchap of this volume, yet the origin of optical emission from TDEs is unknown. As reviewed in the \emischap of this volume, the emission could be powered by an inner accretion disk or by shocks of the intersecting debris streams. A related open question concerns the observed fluence, which is a few orders of magnitude lower than the expected energy released when a star gets disrupted and accreted by a massive black hole \citep{Piran:2015a}. 

The most recent models for the optical emission from TDEs all have been tailored to explain the current set of optical/X-ray properties of TDEs (e.g., peak luminosity, light curve shape, emission line widths). As such, finding new TDEs alone is not sufficient to discriminate between these models. To make progress towards understanding the emission mechanism of TDEs, probing a broader set of observables is key. 

This brings us to the topic of this chapter: we will review a number of observational methods that probe beyond a single-wavelength light curve or spectrum. Each of these methods has provided valuable information about the TDE emission mechanism. However, given the requirements of high signal-to-noise or long monitoring campaigns, these observables have been measured for only a handful of events.  

The techniques discussed here span a very wide range of physical scales. We start with the largest scale of $\sim 1$ light year, as probed by echoes of the TDE emission from dust and gas, measured using infrared imaging (section~\ref{sec:ir}) and coronal lines (section~\ref{sec:cl_lines}), respectively. Next, we consider correlations between the emission at difference wavelengths that have been reported for the TDE ASASSN-14li \citep{Holoien:2016b}: a cross-correlation between the X-ray emission and radio emission (section~\ref{sec:radiox_cc}), which implies a coupling on scales of $\sim 10^{16}$~cm, and a correlation between the X-ray light curve and the UV light curves (section~\ref{sec:uvx_cc}), probing scales of $\sim 10^{14}$~cm. Finally, we consider two results from X-ray timing analysis that probes the innermost parts of the newly formed TDE accretion disks: the detection of quasi period oscillations (QPOs) for some TDEs (section~\ref{sec:qpo}) and relativistic reverberation from the jetted TDE Swift~J1644+57 (section~\ref{sec:xreverberation}). We close with a brief discussion (section~\ref{sec:discussion}).

\section{Infrared imaging: dust echoes}\label{sec:ir}
The reprocessing of TDE light by dust can yield an infrared (IR) echo that probes the environment of TDEs on a scale of $\sim 0.1$~pc. This subject has seen a flurry of activity in recent years. The possibility of dust reprocessing was explored for the first time by \citet{Komossa2009}, 
who presented a {\it Spitzer} observation (Fig.~\ref{fig:SDSSJ0952-Spitzer}) of the coronal-line selected TDE  SDSS\,J0952+2143 (events selected by coronal lines are discussed in section~\ref{sec:cl_lines}). The paper on dust echoes by \citet*{Lu16} made a prediction that infrared emission from a few on-going TDEs could be detectable with imaging observations. Indeed a few months later, three papers that presented observations of dust echoes for different TDEs all appeared in the same week \citep{van-Velzen:2016a,Jiang16,Dou16}---although, as evidenced by \citet{vanVelzen_Spitzer15}, some of these echoes were detected and studied prior to the forecast of \citet{Lu16}.

In the following three subsections we will review these advances in more detail. First, we give a brief overview of the theoretical background of dust echoes. Next, we present the observations of dust echoes using follow-up observations of TDEs. And third, we look at the possibility to discover new TDEs using IR observations of echoes. 

\subsection{Theory of dust reverberation} 
Reverberation by dust has long been used as a tool to study the dust around active galactic nuclei (AGN) \citep[e.g.,][]{Barvainis87} and supernovae and supernovae remnants \citep[e.g.,][]{Dwek83}. The concept is relatively simple. Dust will efficiently absorb  optical/UV/X-ray photons, this process heats the dust to a maximum temperature, which yields isotropic emission that peaks at a few micron. The time delay between the fluctuations in emission from the accretion disk and the IR emission yields the inner radius of the dust region \citep[for an example of the application of this technique in AGN, see][]{Koshida14}. In a TDE, the black hole was not actively accreting, which will modify our expectation of the dust reprocessing signal. The lack of accretion disk emission in the years prior to the TDE means the dust can survive closer to the black hole. But at the same time, the lack of gas supply to the galaxy center implies we could expect a very low dust covering factor and thus a weak reverberation signal compared to active galaxies. 

To estimate relevant timescale of a dust echo we have to equate the absorbed flux at radius $R$ from the black hole to the emission by the dust particles. The latter is relatively well-constrained, dust absorption is efficient across the wavelength range where TDEs emit most of their energy. However the dust emission is more uncertain because it depends on the model for the dust particles, which can include grain composition, size, and the shape of the emission spectrum as free parameters. 

Fortunately, the uncertainty on the grain properties is reduced by the extreme conditions of a TDE, which reduce the type of dust that we expect to be visible. The grains that can survive closest to the black hole dominate the IR echoes because they radiate the received energy in the shortest amount of time. Furthermore, if the dust emission efficiency scales with the square of the dust size \citep{Draine84}, we expect the largest grains to dominate the reprocessing emission. Hence most authors have normalized their predictions for TDE dust emission to a single grain size, using $a\approx 0.1$~$\mu$m, the typical maximum dust size for the Milky Way \citep{Weingartner01}. 

After restricting to ``graphite-like" dust particles, because these have the highest sublimation temperature and should thus dominate the reprocessing signal,  \citet{van-Velzen:2016a} arrive at the following expression for the sublimation radius 
\begin{equation}\label{eq:Rsub1}
    R_{\rm sub} = 0.15 \left(\frac{L_{45}}{a_{0.1}^2 T_{1850}^{5.8}}\right)^{1/2}~{\rm pc} \quad. 
\end{equation}
Here $L_{45}$ is the bolometric luminosity of the TDE in units of $10^{45}$~erg\,s$^{-1}$, $a_{0.1}$ is the grain size in units of 0.1~$\mu$m, and $T_{1850}$ is the temperature of the dust, normalized to the expected sublimation temperature of 1850~K.  Eq.~\ref{eq:Rsub1} is derived with the common assumption that dust absorption is 100\% efficient across the relevant wavelength range ($Q_{\rm UV}=1$). 
If we use a dust model that approximates a two-component medium consisting of both graphite and silicate grains \citep{Waxman00,Lu16} the sublimation radius is 
\begin{equation}\label{eq:Rsub2}
    R_{\rm sub} = 0.12 \left(\frac{L_{45} (1+0.1 a_{0.1} T_{2300})}{a_{0.1} T_{2300}^5}\right)^{1/2}~{\rm pc} \quad, 
\end{equation}
with $T_{2300}$ the dust temperature normalized to 2300~K.

From both Eq.~\ref{eq:Rsub1} and Eq.~\ref{eq:Rsub2} we find a typical size scale of 0.1~pc for the emitting region, which implies a timescale of a few months for the reprocessing signal. Because this timescale can be measured from the IR light curve (Fig.~\ref{fig:PTF09ge}), solving $R_{\rm sub}$ for $L_{45}$ yields an estimate of the bolometric luminosity of the TDE. This approach only works if the observed IR emission is dominated by a single shell that emits at the sublimation temperature.  If the dust echo contains emission from larger scales (e.g., because the galaxy is devoid of dust at $R<R_{\rm sub}$), a measurement of the dust temperature is also needed to estimate $L_{\rm bol}$ \citep[e.g.,][]{Kool20}. 

While the bolometric luminosity is estimated from the {\it shape} of the reprocessing light curve, the covering factor ($\Omega_d$) follows from the {\it amplitude} of the echo. We define $\Omega_d\equiv E_{\rm bol} / E_{\rm dust}$, with $E_{\rm dust}$ the total energy radiated by the dust and $E_{\rm bol}$ the total radiated energy of the TDE (integrated over the wavelength where dust absorption is efficient). 

\begin{table}[bt]
    \caption{Events with transient infrared emission.}
    \begin{centering}
    \begin{tabular}{l c c l}
    \hline
    Name        &  {Host galaxy type} &  Peak IR luminosity$^{*}$    & Reference \\
                &       &  (erg\,s$^{-1}$)          & \\
    \hline\\[-7pt]
    PTF-09ge            & Quiescent   &  $10^{41.8}$ & 1 \\
    PTF-09axc           & Quiescent (E+A)  &  $10^{42.6}$ & 1 \\ 
    ASASSN-14li         & Quiescent (E+A)  &  $10^{41.6}$ & 2 \\ 
    AT2019dsg           & Quiescent   &  $10^{43.0}$ & 3$^{\dagger}$  \\
    {NGC 5092}          & Quiescent & $10^{42.6}$& 4 \\
    IRAS F01004-2237 & ULIRG     & $10^{44.3}$ & 5 \\ 
    PS16dtm         & AGN (Type 1)      & $10^{43.0}$ & 6 \\
    {PS1-10adi}       & AGN (Type 1)    & $10^{44.0}$ & 7,8 \\
    SDSS J1657+2345 & AGN (Type 2)     & $10^{43.9}$ & 9 \\
    SDSS J0227-0420 & AGN (Type 1)     & $10^{43.6}$ & 10 \\
    Arp 299-B\,AT1  & LIRG     & $10^{43.8}$ & 11 \\
    {AT2017gbl}       & LIRG & $10^{43.2}$ & 12 \\ 
    {MCG-02-04-026} & AGN (Type 1) & $10^{43.2}$ & 13 \\
    SDSS J0952+2143      & {Quiescent (HII-type)} & $10^{43.9}$ & 14,\,15 \\
    \hline \\[-5pt]
    \end{tabular}
    \end{centering}
    \small
    \\Notes --- {This table includes events published up to 2020, for more recent IR-detection  from optical TDEs see \citet{Jiang21}, which appeared while this article was under review.} \\ 
    $^{*}$Peak observed luminosity ($\nu L_\nu$) based on host-subtracted WISE data (except for SDSS J0952+2143 the peak flux of which  is based on {\it Spitzer} photometry).  \\
     $^{\dagger}$The peak IR luminosity of AT2019dsg is based the host-subtracted neoWISE magnitude of $W1=13.1$ (Vega). \\ 
    References: 
    1 \citep{van-Velzen:2016a}, 
    2 \citep{Jiang16}, 
    3 \citep{Stein20},  
    4 \citep{Li20_NGC5092},
    5 \citep{Dou17}, 
    6 \citep{Jiang17}, 
    7 \citep{Kankare17}, 
    8 \citep{Jiang19}, 
    9 \citep{Yang19}, 
    10 \citep{Liu20}, 
    11 \citep{Mattila18},
    12 \citep{Kool20}, 
    13 \citep{Sun20}
    14 \citep{Komossa2009}, 
    15 \citep{Dou16}.
    \label{tab:ir}
\end{table}


\subsection{Infrared echoes detected in follow-up observations} 

\begin{figure}
\includegraphics[clip, trim=1.7cm 1.0cm 1.3cm 14.5cm, width=\columnwidth]{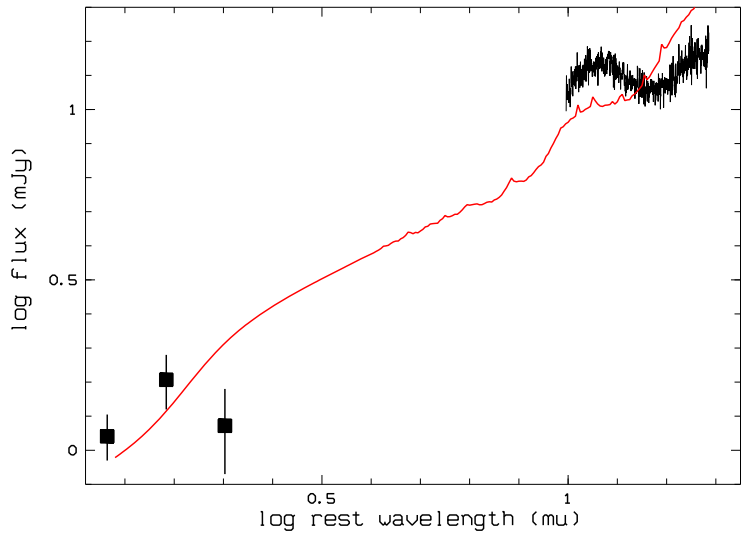} \caption{IR SED of SDSS\,J0952+2143 observed with Spitzer (black solid line) and 2MASS (pre-flare; filled squares), reproduced from \cite{Komossa2009}. {The plot emphasizes the large Spitzer IR luminosity at outburst in comparison to pre-outburst NIR data, and the characteristic Spitzer spectral shape, measured for the first time in a TDE.} Overplotted is the mean Spitzer IR SED of accreting extragalactic sources (solid red line; \citealt{Netzer07}. Note a conversion error in \citet{Komossa2009} when plotting that comparison SED (their Figs. 7 and 8), which led to an incorrect presentation of the shape of that SED, corrected by \citet{Komossa10}. } 
\label{fig:SDSSJ0952-Spitzer}
\end{figure}

The first TDE with IR follow-up observations was SDSS\,J0952+2143 \citep{Komossa2008,Komossa2009}, which was selected based on variable coronal line emission and variable optical, NIR and UV continuum emission (this class of events is discussed in full detail in the next section). 
This source was observed once with the {\it Spitzer} IRS spectrograph at 10--20 $\mu$m (Figure~\ref{fig:SDSSJ0952-Spitzer}), yielding three main results  \citep{Komossa2009}. First, the detection of a large  MIR luminosity of $L_{\rm 10-20\mu m} = 3.5 \times 10^{43}$ erg\,s$^{-1}$ in the {\it Spitzer} band which implies a large X-ray luminosity (not observed directly) and efficient reprocessing. Second, a blackbody fit to the IR spectrum gave an estimate of the radius of the IR emission site $r\approx 0.5$\,pc.  Third, the characteristic emission features from PAHs and silicate dust were detected in the spectrum. 

The single {\it Spitzer} observation by \citet{Komossa2009} was carried out about 4 years after the start of the optical flare \citep{Palaversa16}. The transient nature of the IR at emission was confirmed later via WISE observations at 12$\mu m$ (the WISE W3 band), obtained 1.9 year after the {\it Spitzer} spectrum, showing that the flux at this wavelength has decreased by a factor of $1.8$ \citep{Dou16}. 

Infrared photometric data of optically discovered TDEs are often obtained from the Wide-field Infrared Space Explorer \citep[WISE;][]{Wright10} which has been systematically scanning the whole sky in $3.4$ and $4.6\mu$m since 2010. The primary (cryogenic, 4 band) and post-cryogenic {\it WISE} surveys ran from Jan 2010 to Feb 2011. Due to lack of funding, the {\it WISE} space telescope was placed in hibernation between 2011 to 2013. 
On December 13, 2013, NASA reactivated the {\it WISE} telescope and started an all sky survey called Near-Earth Object WISE Reactivation mission \citep[NEOWSIE-R;][]{Mainzer14}. {NEOWISE-R} is still operating since December 2013. It scans the most part of the sky at $3.4$ and $4.6\mu$m every half year, with $10-20$ single exposures in each visit.   

Based on the {\it WISE} archive photometry, several studies have examined the mid-IR (MIR) light curves (LC) of TDEs identified in optical imaging surveys with spectroscopic follow-up observations, see the \optchap of this volume. These include ASASSAN-14li \citep{Jiang16}, PTF09ge, and PTF09axc \citep{van-Velzen:2016a}. Several optical flares in active galaxies {that have been interpreted as TDEs} are also found to show mid-IR flares, see Table~\ref{tab:ir}. \citet{Dou16} presented WISE detections of four SDSS galaxies with variable coronal emission lines (see section~\ref{sec:cl_lines} of this chapter). It is interesting to note that the MIR LCs of the spectroscopically-selected TDEs, {if detected in the first place,} all have small amplitude variations with $\Delta m \sim 0.1$ magnitude. For ASASSN-14li and PTF-09ge, the peak MIR LC corresponds to infrared luminosity $L_{\rm IR} \sim 10^{41}$\,erg\,s$^{-1}$ and $L_{\rm IR} \sim 10^{42}$\,erg\,s$^{-1}$, respectively. {As shown in Table~\ref{tab:ir}, this luminosity is between one and two an orders of magnitude lower than the peak IR luminosity of the TDE candidates in active galaxies.}

\begin{figure}
\centering
\includegraphics[width=300pt]{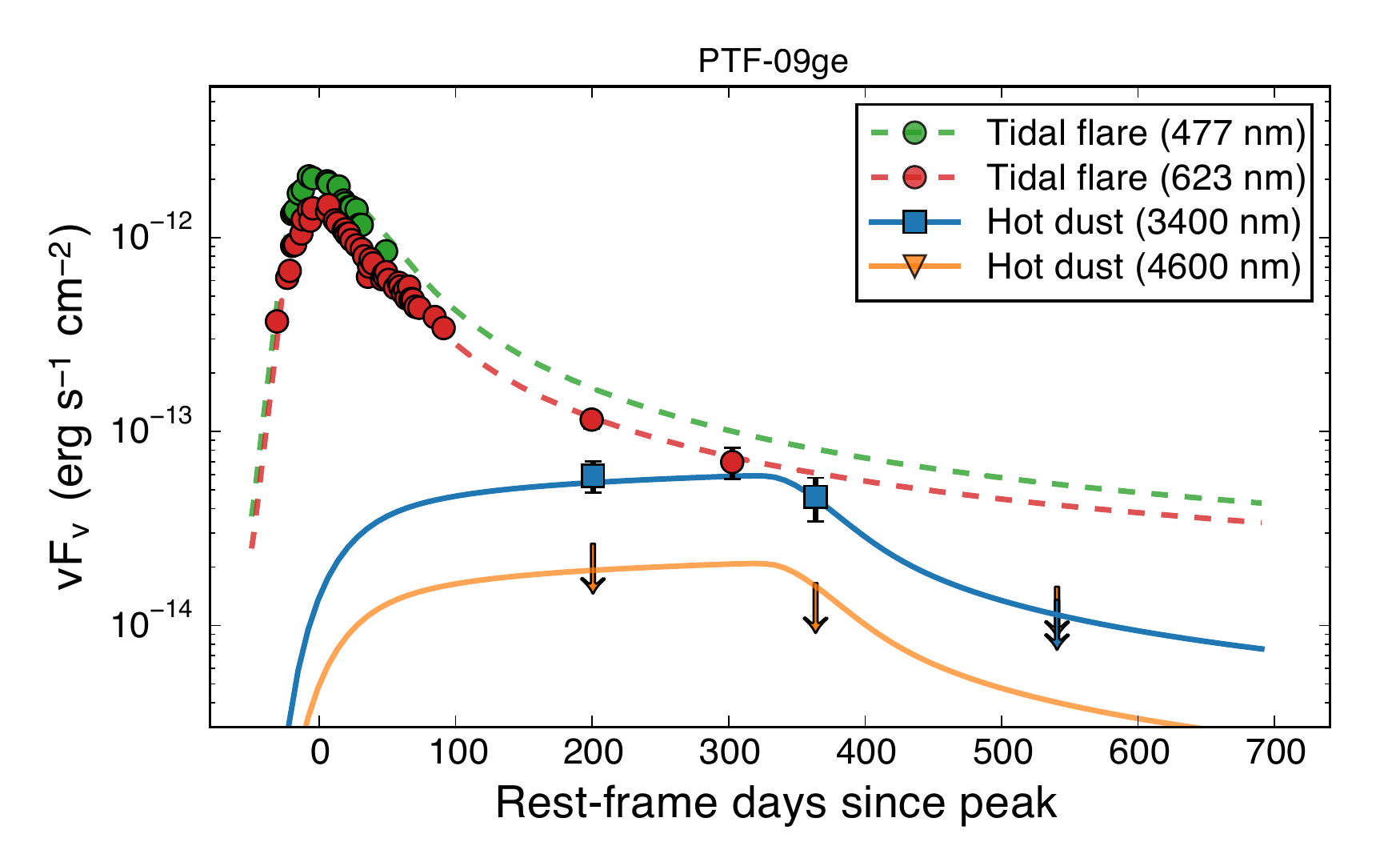}
\caption{The optical \citep{Arcavi:2014a} and infrared  \citep{van-Velzen:2016a} light curve of the TDE PTF-09ge. The solid lines show the dust reverbation light curve, which is derived from the optical light curve using only the radius of the dust shell and the covering factor as free parameters. The former sets the duration of the event, while the latter affects only the overall normalization. The best-fit parameters are $R=0.15$~pc and $f_{\rm dust}=0.01$. Figure reproduced from \citet{van-Velzen:2016a}.}
\label{fig:PTF09ge}       
\end{figure}

The peak phases of the MIR LCs display time delays relative to the optical LCs, with $\Delta t \sim 100 - 200$\,days. This time delay translates to the radius of dust medium of about $0.1$\,pc, somewhat larger than the dust sublimation radius of 0.04\,pc as derived from $L_{UV}$. In general, these numbers are consistent within the context of MIR LCs as the result of the dust echo of the UV/optical flares. Three additional parameters -- the covering factor, dust mass and geometry, and the flare bolometric luminosity (via Eq.~\ref{eq:Rsub1}) -- can be estimated from the MIR LCs, although with considerable uncertainty because dust temperatures are not well constraint with only one color ($[3.4 - 4.6]\mu$m). 
For PTF-09ge (Figure~\ref{fig:PTF09ge}), \citet{van-Velzen:2016a} marginalized over the uncertainty on the dust temperature and included the measurement uncertainty on $R_{\rm sub}$ to find $\log_{10}(L_{\rm bol}) = 44.9^{+0.1}_{-0.6}$ and $\log_{10} \Omega_d = -2.0 \pm 0.2$.  

Better spectroscopic coverage is one area of improvement for future near-IR follow-up observations. And with well-sampled IR light curves, we can obtain a better handle of the geometry of the dust, allowing us to differentiate between a torus model \citep{Dou16,Sun20} and a polar or spherical dust geometry \citep{van-Velzen:2016a,Mattila18,Kool20}.   

The lower average IR luminosity of TDEs in quiescent galaxies compared to events selected by their variable coronal emission lines (ECLEs, see section 3) suggests that the covering factor and the dust mass that participates in the reprocessing is larger for the ECLEs. Intriguingly, the first {\it WISE} detections of the MIR flare in SDSS\,J0952+214 is 3--5 years after the changes of coronal emission lines. Together with the high IR luminosity, this long time scale of the variation could be an indication that these do not belong to the same class as the TDEs selected based on optical photometric and follow-up spectroscopy. We discuss the ambiguity between changing-look AGNs and TDEs, and the possible ways to separate them in section~\ref{sec:IRcompare} below.   

Infrared echo offers a unique tool to probe the dusty circumnuclear medium at close distances to the central supermassive black holes. One interesting example is PS16dtm, which could be a TDE  in a Narrow Line Seyfert 1 galaxy \citep{Blanchard:2017a}--- however see \citet{Moriya17} for a different interpretation. The MIR LC of PS16dtm appears to rise about $11$\,days before the first epoch of optical detection \citep{Jiang17}. This {can be explained by a pre-existing dusty medium} with a size of 11 light days ($2.8\times10^{16}$\,cm), with a sufficient high covering factor, blocking the initial optical/UV light. This size is similar to the dust sublimation radius inferred from the quiescent AGN with $L_{\rm bol}\approx 5\times10^{42}$\,erg\,s$^{-1}$. After the TDE flare, the high flux of UV/X-ray photons destroys the dust grains and enlarges the inner radius of the dust free cavity to $\sim 70$ light days ($2\times10^{17}$\,cm). Additional Fe elements previously locked up by the dust grains are now released into the ionized gas medium, potentially explaining the abundant FeII emission lines observed in the follow-up spectra by \citet{Blanchard:2017a}. 

{Infrared echoes appear to be ubiquitous to the class of highly energetic transients from the centers of active galaxies \citep{Kankare17}. An example is PS1-10adi \citep{Kankare17}, which has a very high MIR luminosity of $\sim 10^{44}$\,erg\,s$^{-1}$ \citep{Jiang19}.

Infrared flares interpreted as dust reprocessing signals of energetic transients have also been found in star-forming and dusty luminous infrared galaxies (LIRGs; the infrared emission in these galaxies is thought to be mainly powered by star formation \citealt{Genzel2020}). The best-studied case is Arp 299-B\,AT1, which was observed with high spatial resolution radio observations combined with near- and mid-IR observations covering a period of over 10 years \citep{Mattila18}. Based on this excellent data, a TDE is the preferred explanation proposed by \citet{Mattila18}. 

Due to the small sample size of LIRGs that have been monitored for IR flares, this interpretation would imply an elevated rate of TDEs for this subset of galaxies. Likewise, a TDE interpretation of the transients found by \citet{Tadhunter17} in a small sample of ultra-luminous infrared galaxies (ULIRGs) would imply a factor $\sim 100$ enhancement to the TDE rate for this galaxy class \citep{Stone18a}. } 


\subsection{Infrared flares from quiescent and active galaxies: a tool to find dust-obscured TDEs?}\label{sec:IRcompare}  
TDE studies using optical all-sky transient surveys face several important questions. First, tidal disruption of stars by central black-holes can in principle also occur in galaxies with active galactic nuclei (AGN) and dusty infrared luminous star-forming galaxies. Second, how do we distinguish Changing-Look (CL) AGNs from TDEs? While the underlying physics has similarities, these types of transients may probe different scales (time, size and mass) of accretion phenomena. 

Some TDE candidates have been found in Seyferts and dusty LIRG/ULIRGs, {as listed in  see Table~\ref{tab:ir}}. These studies confront a common question---could these flares be due to AGN changing their states, i.e., CL AGNs? 

The observational ambiguities between CL AGNs and TDEs are abundant---see the \impostchap of this volume. Separation often requires multi-wavelength data over a long time scale of several years, as has been obtained for the systems discussed in section 3. Particularly, the properties of a ``turn-on'' AGNs, transitioning from a quiescent to an active accreting state, can be easily confused with that of a TDE, as shown by \citet{Yan19}.
Using an extensive dataset of {\it Swift} UV, optical, mid-IR LCs and spectroscopy over 1200\,days, this study made comprehensive comparisons between CL AGN, TDE and unusual SN\,IIn and showed that the flare in SDSS1115+0544 is a rapid ``turn-on" of a type 1 AGN, changing from its earlier quiescent state. \citet{Yan19} highlighted that having only LCs or variable spectra, including high ionization coronal lines, is not sufficient to make conclusive classifications (see also the detailed discussion by \citealt{Komossa2008}). A potential solution could be collecting larger samples of the various classes of TDEs, including events in AGNs and LIRGs/ULIRGs.

Potentially, using infrared variability should offer an effective means to identify TDE candidates in dusty galaxies. Such systematic searches using {\it WISE} mid-IR time series photometry are being carried out and below we discuss an example of this approach.

Using only the {\it WISE} data between December 2009 and February 2011, \citet{Wang18} searched for potential TDE candidates among SDSS galaxies with spectra which present no indications of AGN. Only based on their mid-IR LCs, they found a total of 14 TDE candidates from quiescent SDSS galaxies at $z<0.22$ with estimated blackhole masses of $10^{7-8}M_\odot$. The estimated  rate is $\sim 10^{-4}$\,galaxy$^{-1}$\,yr$^{-1}$ with large uncertainties. As concluded by these authors, these 14 candidates could contain a good fraction of CL AGNs which show short time scale episodic variability. 

The second ongoing search is done by the same group but with the entire {\it WISE} data between December 2009 and 2018 (see \citealt{Yan19} for a brief description of the full survey). {The systematic search for infrared flares was done among all SDSS spectroscopic confirmed galaxies at $z < 0.35$. A total of 137 galaxies were selected with flare amplitudes $\ge 0.5$\,magnitude in at least one WISE band \citep{Jiang2020}. This sample does not have complete spectroscopic classification. It contains a mixture of various types, including a small fraction of supernovae and radio loud AGNs. \citet{Jiang2020} speculates that a large fraction of this sample are possible dust echoes of TDEs and CL AGNs.} 
One example with extensive spectroscopy is  SDSS1115+0544, a rapid ``turn-on" type-1 AGN \citep{Yan19}.

{Radio follow-up observations of 16 {\it WISE}-selected flares by \citet{Dai20} have yielded a high detected fraction of 75\% (for a typical 5$\sigma$ upper limit 60$\mu$Jy at 5.5~GHz). Based on pre-flare  SDSS spectroscopy, about half of targets in this sample can be classified as AGN, while most of the non-AGN show evidence for active star formation. A radio detection rate of 75\% is much higher than the fraction of AGN or TDEs that are considered to be ``radio-loud", see the \radiochap of this volume.
Based on this high detection rate and similarities with other radio-detected IR flares \citep{Mattila18,Kool20}, one could conclude that IR-selection yields a subclass of TDEs that are often accompanied by powerful jets, with total energies of $\sim 10^{52}$~erg. Such powerful jetted TDEs have been observed before (see \citealt{Bloom11,Zauderer11}, and the \gammachap of this book), although the estimated volumetric rate of these events, $\sim 1$\,Gpc$^{-1}$yr$^{-1}$ or $\sim 10^{-7}$\,galaxy$^{-1}$\,yr$^{-1}$ \citep{vanVelzen18_ngVLA}, is three orders of magnitude lower than typical rate of WISE-selected flares from galaxies with SDSS spectra ($\sim 10^{-4}$\,galaxy$^{-1}$\,yr$^{-1}$; \citealt{Wang18}).  Alternatively,  one could  conclude that a selection of IR flares in WISE data yields a sample of active black holes in a special phase of their evolution, during which a large change in the accretion rate triggers or rekindles jet activity, leading to a higher detection rate of radio emission. }

\section{Optical spectroscopy: emission-line echoes }\label{sec:cl_lines}

 \subsection{Introduction}


If TDEs occur in a gas-rich environment, their luminous electromagnetic radiation will photoionize any surrounding gas and will be reprocessed into emission lines.  These are then very useful probes of the physical conditions and kinematics of the gaseous material in the SMBH vicinity, including the disrupted star itself. 

Photoionization by the luminous TDE radiation, which often peaks in the EUV or soft X-rays, will then not only produce Hydrogen Balmer lines, but in particular also high-ionization lines like HeII$\lambda$4686 and coronal lines. The ions which produce these lines have their ionization potentials (IPs) in the EUV to soft X-ray domain. Several such events have been identified in the last decade based on  Sloan Digital Sky Survey (SDSS) spectroscopy and multi-wavelength follow-up observations 
\citep{Komossa2008,Komossa2009,Wang11,Wang12,Yang:2013a,Palaversa16,Dou16}
and these are discussed below. 

\begin{figure}
\centering
{\includegraphics[height=7.6cm]{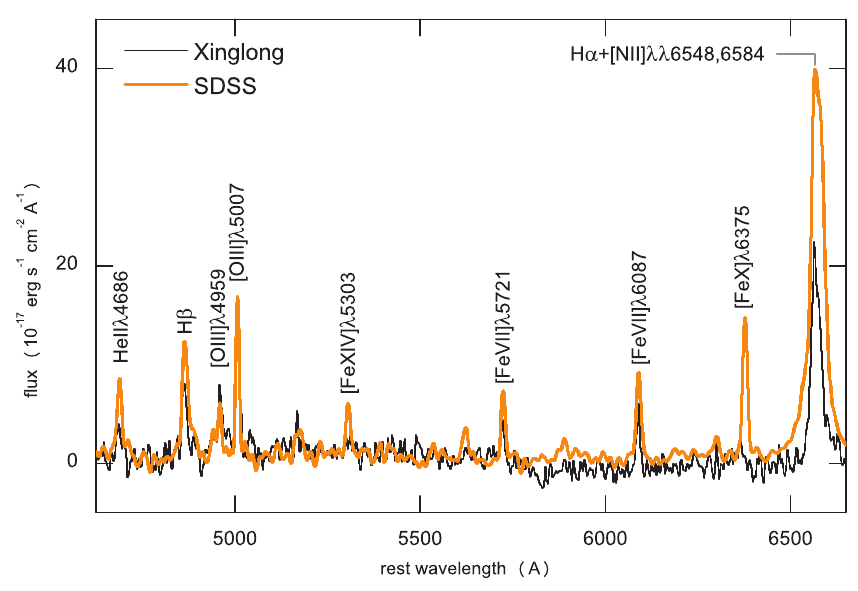}}
\caption{Optical SDSS spectrum of SDSS\,J0952+2143 during the high-state (orange) (2005) with several strong iron coronal lines {and other transitions} marked. A zoom on the H$\alpha$ profile (shown above in degraded spectral resolution, to match the same resolution as the Xinglong spectrum) is presented in Figure \ref{fig:J0952-ecle-balmerforns} in high resolution. The first optical follow-up spectrum (2007, taken with the Xinglong telescope, plotted in black) revealed the strong fading of the highest-ionization lines \citep{Komossa2008}.
}
\label{fig:J0952-ecle-broadband}       
\end{figure}

The best-studied of these events, SDSS\,J0952+2143, (Table~\ref{tab:cl}) was first noticed because of its unusual Balmer and HeII emission profiles and especially the exceptional strength of its high-ionization iron lines relative to [OIII]$\lambda$5007 \citep{Komossa2008} which were, however, not permanent, but faded away dramatically on a timescale of years \citep{Komossa2009,Yang:2013a},  see Figure~\ref{fig:J0952-ecle-broadband}.  
A systematic search in SDSS for similar events \citep{Wang12} let to the  identification of six more, of which three \citep{Yang:2013a} showed strongly fading coronal lines (Table~\ref{tab:cl}). Key selection criterion was the presence of bright coronal line emission of [FeX]-[FeXIV] in SDSS spectra, which had dropped significantly in intensity in dedicated follow-up optical spectroscopy.

\begin{table}[tb]
\begin{centering}
\caption{Extreme coronal line emitters}
    \begin{tabular}{l c l}
    \hline
    Full (short) name & Redshift  & First reference \\ \\[-7pt]
    \hline
    SDSS\,J095209.56+214313.3 (J0952+2143) & 0.0789 & \citealt{Komossa2008} \\
    SDSS\,J074820.67+471214.3 (J0748+4712)  & 0.0615 & \citealt{Wang11}        \\
    SDSS\,J134244.42+053056.1 (J1342+0530) & 0.0366 & \citealt{Wang12}        \\
    SDSS\,J135001.49+291609.7 (J1350+2916) & 0.0777 & \citealt{Wang12}        \\
    \hline
    \end{tabular}
    \label{tab:cl}
\end{centering}
\end{table}

Among these extreme coronal line emitters (ECLEs hereafter), SDSS \\ J0952+2143 at redshift $z=0.079$ is the one with the most luminous coronal lines, and the densest multi-wavelength coverage both near the high-state and in subsequent years \citep{Komossa2008,Komossa2009} to which \citet{Palaversa16} added a long-term optical light curve. Follow-up spectroscopy three years after the SDSS high-state spectrum revealed that the Balmer and coronal lines had strongly declined (Figure~\ref{fig:J0952-ecle-broadband}). The fact that the outburst extended into the high-energy regime was indirectly inferred  from the presence of the luminous, fading coronal lines, while at lower energies continuum variability was directly detected in the NIR, optical and UV \citep{Komossa2008}, followed by a continuous fading of the long-term optical light curve \citep{Palaversa16}. 
  
The host galaxy of SDSS\,J0952+2143 was classified as inactive, as no signs of permanent AGN activity were identified; based on line-ratios, NIR colours, and faint X-rays in low-state (\citealt{Komossa2008,Komossa2009}; our Sect. 3.5). Therefore, the observations are consistent with a TDE which happened in a gas-rich environment.

In addition to the ECLs, fading broad Balmer and He line emission dominate the optical spectrum, with asymmetric profiles, further covered in the \optchap of this book.

\subsection{Iron coronal line diagnostics} 


The presence of the luminous and transient high-ionization lines in optical spectra of the galaxies immediately implied that the sources underwent luminous electromagnetic flares which extended into the soft X-ray regime  -- not directly observed at all wavebands, but indirectly inferred from the presence of the fading emission lines, 
as the ions which produce these lines have their ionization potentials (IPs) in the EUV to soft X-ray regime. The iron lines, with initial luminosities up to $L_{\rm [FeVII]-[FeXIV]} > 10^{40}$ erg\,s$^{-1}$ in SDSS\,J0952+2143, faded away on the timescale of years, and the degree of ionization changed from higher ([FeX] to [FeXIV]) to lower ([FeVII]), and up to the complete lack of coronal lines 5-8 years after high-state.  

Further, the iron lines are known to be sensitive diagnostics of the physical conditions in the line emitting gas. Individual line ratios enable us to distinguish whether the main ionization mechanism is collisional ionization or photoionization, while other line ratios are density and temperature sensitive \citep[e.g.,][]{NussbaumerStorey82,KeenanNorrington87,Dere09}. Based on these diagnostics, gas temperatures of the order of $(1.5-3)\times 10^4$~K were inferred, consistent with a photoionization origin of the emission lines,  and gas densities around $n = 10^7$ cm$^{-3}$ \citep{Komossa2009,Wang12}.



\subsection{Oxygen line diagnostics}

In SDSS\,J0952+2143, the intensity ratio of [OIII]4363/[OIII]5007 is exceptionally high \citep{Komossa2008}. [OIII]4363/[OIII]5007 $\simeq$ 0.2--0.3 is significantly above the value observed in AGN and above photoionization predictions for a large parameter range \citep{KomossaSchulz97,Groves04}. The observed value implies a high-density regime, where the ratio [OIII]4363/[OIII]5007 no longer only depends on temperature, but becomes a density diagnostic \citep{Osterbrock89,DopitaSutherland03}. The observed value implies $n = 10^7$ cm$^{-3}$ (for $T=(2-5) \times 10^{4}$ K), and the [OIII]4363 emission therefore arises in a similar region as the ECL emission \citep{Komossa2009}. 

Two other ECLEs, SDSS\,J\,0748+4712 and SDSS\,J1342+0530, show an unusual ratio [OIII]5007/[OIII]4959 $\approx 2$ in late-time spectra, strongly deviating from the canonical theoretical value of 3 observed in all AGN, and interpreted as sign of high density and high optical depth. Unless the metallicity is exceptionally  high ($Z > 10$), the data imply that the [OIII] emission region is optically thick to electron scattering \citep{Yang:2013a}. 

\subsection{Balmer-line diagnostics: Double-peaked horns in H$\alpha$ and H$\beta$} 

\begin{figure}
\centering
{\includegraphics[height=6.3cm]{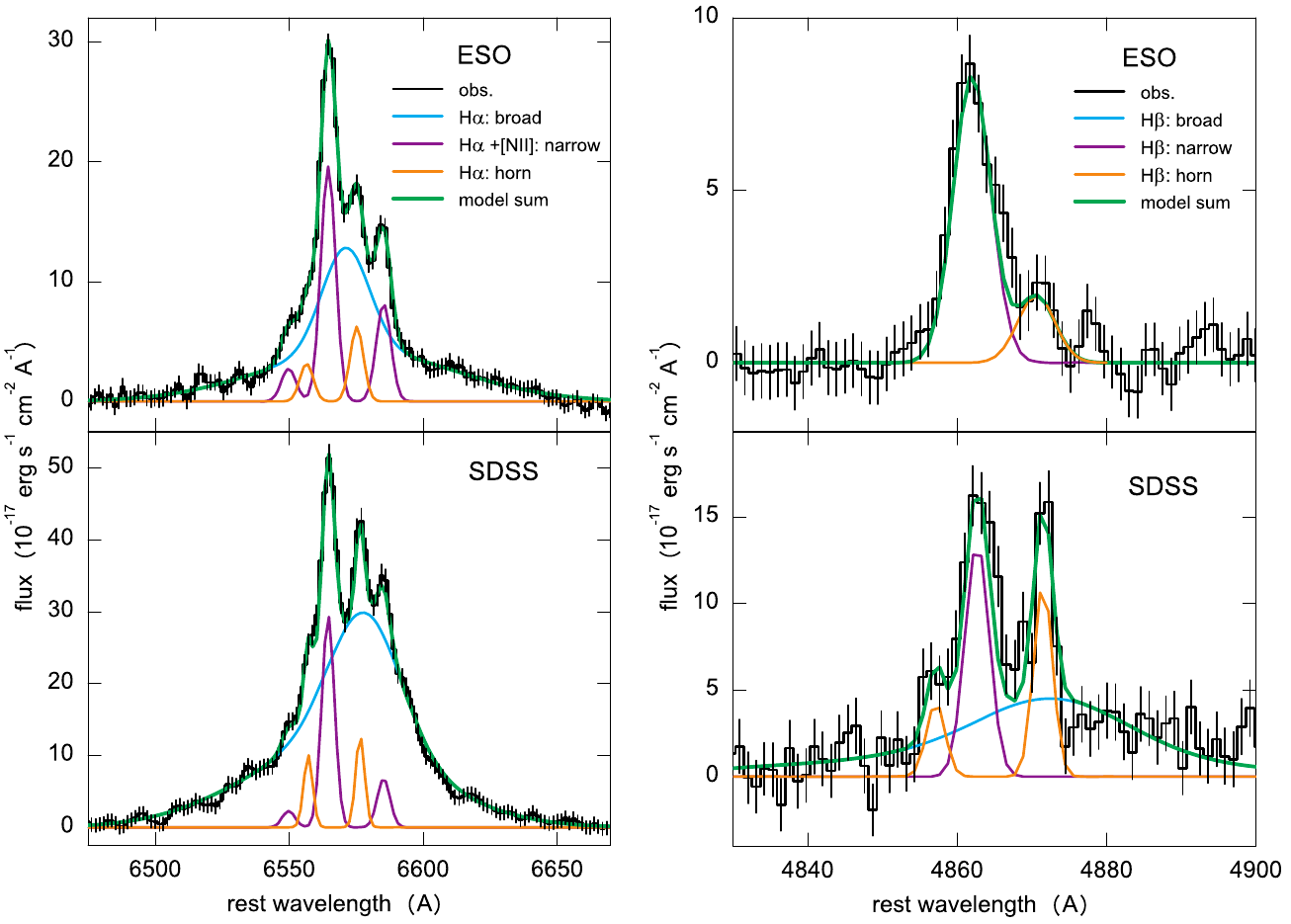}}
\caption{Hydrogen emission-line profiles of  SDSS\,J0952+2143 (H$\alpha$ left, H$\beta$ right) taken at SDSS high-state in 2005, and with the ESO NTT in 2008. In addition to an asymmetric, broad base (blue), two narrow horns (orange) are clearly present. Their width is unresolved. Adapted from \citet{Komossa2009}.}
\label{fig:J0952-ecle-balmerforns}       
\end{figure}

A narrow double-peaked structure in the Balmer lines has been detected in the optical spectra of SDSS\,J0952+2143 \citep{Komossa2008}, not seen in any other of the ECLEs \citep{Wang12}, or any other source we are aware of.
At high-state, the H$\alpha$ and H$\beta$ lines of SDSS\,J0952+2143 show a multi-component profile, which can be decomposed into at least three distinct components (Figure \ref{fig:J0952-ecle-balmerforns}): 
(1) a broad component of FWHM(H$\alpha$)=1930 km/s, asymmetric and redshifted by 570 km/s, (2) a narrow component at the same redshift as other narrow  forbidden emission lines, and (3) two very narrow horns (FWHM $<$ 200 km/s, unresolved), one redshifted and the other blueshifted. While line intensities of the Balmer line components and especially the two narrow horns had faded significantly 3 years after the SDSS high-state, their velocity shifts of $v$=540 km/s (redshifted horn) and $v=-340$ km/s (blueshifted horn) remained remarkably constant \citep{Komossa2009}. 
 
The exact nature of the double-horn emission remains unknown at present. The line shape, i.e., the double-peaked profile, could arise in a ring or disk  geometry or a two-sided outflow. Surprisingly, the two horns do not have a counterpart in any other of the emission lines. While at first glance, this fact could imply high gas density, several of the strong coronal lines have very high critical density, too, and yet they do not show multiple horns.  \citet{Komossa2009} therefore discussed the possibility that these lines form in collisionless shocks. Such shocks produce strong H lines with a two-component profile composed of a narrow component contributed by cold H atoms, and a broad component which arises from H atoms that have undergone charge transfer reactions  with hot protons while other optical forbidden lines are very faint \citep[e.g.,][]{Raymond95,HengMcCray07}. Alternatively the broad underlying Balmer line with its two additional narrow horns may represent emission from the inner TDE accretion disk \citep{Komossa2008}.

\begin{figure}
\centering
{\includegraphics[height=4.6cm]{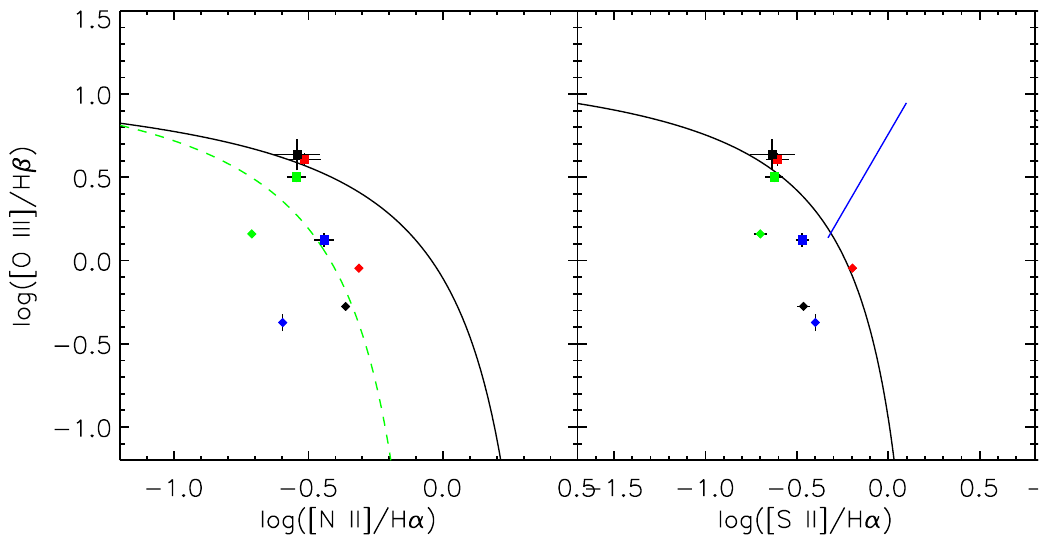}}
\caption{Location of ECLEs in diagnostic diagrams. They are located in the starburst regime (small symbols), and shift towards higher excitation in the later optical spectra (large squares) which exclude more of the outer starburst emission. The black line marks the dividing line between starformation (lower left) and AGN (upper right), while the blue line represents the deviding line between AGN (above the line) and LINERS (below the line). The region between the green dotted line and black line represents the extreme starburst regime (see \citet{Yang:2013a}).  }
\label{fig:ecle-DDs}       
\end{figure}

\subsection{A note on the interpretation of these events}
In summary, four events were observed which showed strong emission-line flaring in response to an optical/UV/X-ray outburst.
These ECLEs reside in sub$L_{*}$ disk galaxies \citep{Wang12}.  A SMBH mass from stellar absorption lines of 7 $\times 10^6$ M$_{\odot}$ was estimated in one case \citep{Komossa2008}.   
Three outburst scenarios have been discussed in the literature.

    The first outburst scenario involved very unusual AGN variability. However, no positive evidence for the presence of AGN has been identified: emission-line ratios are in the starburst (HII) regime of diagnostic diagrams (Fig. \ref{fig:ecle-DDs}), NIR colours imply quiescent galaxies,  and post-flare X-ray emission, only measured in SDSS\,J0952+2143, is below that of classical AGN ($L_{\rm x} = 10^{41}$ erg\,s$^{-1}$)\citep{Komossa2008,Komossa2009,Wang12}.

    Alternatively, a SN origin has been discussed { \citep{Komossa2009,Wang11}.}
    However, the very luminous iron coronal lines,  especially of SDSS\,J0952+2143, do imply luminous high-energy emission never observed in SNe and  the coronal lines are a factor 100 more luminous than the record holder among SNe, making this event very different from any known type of SN.  Further, the longevity of the UV emission, detected 1.8 years after the start of the flare \citep{Palaversa16},  strongly disfavors an SN origin.
    
Finally, the events can be interpreted as TDEs in a gas-rich environment, which is preferred explanation by most authors \citep{Komossa2008,Komossa2009,Wang12,Palaversa16}. In this scenario, the radiation from the temporary accretion disk photoionizes any circumnuclear gaseous material as well as the disrupted star itself.


\section{Cross correlations}\label{sec:crosscorr}
\subsection{Introduction and methodology}
Multi-wavelength photometric reverberation mapping using X-ray, optical, UV, and radio 
time series of actively accreting SMBHs, i.e., AGN, and stellar-mass black holes have 
yielded significant insights into the geometries of structures surrounding these systems 
\citep[e.g.,][]{mchardyngc4593:2018, edelson:2015, marscher3c120:2018, poshakgx339:2010, 
obrien:2002, chatter09:3C120}. The underlying method in these studies is to first check 
for similarity in variability between two light curves derived from two distinct wavebands, and 
then measure time lags, if any, between them. The hypothesis is that if two light curves 
from two distinct wavebands vary in the same manner, then they are likely driven by a 
common physical mechanism. 

Under the assumption that the emission from the reference waveband 
comes from very near to the black hole, any correlations and subsequently time lags 
give a measure of how far the reverberating regions are with respect to the black hole. 
The direction (leading vs lagging the emission from near the black hole) and the 
magnitudes of the time leads/lags vs wavelength essentially give a 1-dimensional map of 
the geometry of the system under consideration \citep[e.g., Fig. 6 of ][]{cackett:2018}. 
Thus, reverberation mapping analysis provides an invaluable tool to probe spatial scales 
which cannot be resolved with our current telescopes. For accreting stellar-mass black 
hole binaries the relevant timescales are between a fraction of a second to a few tens 
of seconds \citep[e.g.,][]{poshakgx339:2010, obrien:2002} while for AGN they are 
anywhere between a couple of minutes to a couple of months \citep[e.g.,][]{edelson:2015}.

One of the main statistical tools used in reverberation mapping analyses is the so-called 
cross-correlation function (or CCF). Mathematically, this defines the amount of 
similarity between two time series as a function of time offset (time lag/lead) between them. 
CCFs are usually normalized such that a value of unity means a perfect match between the two input time series while a 
value of negative unity implies they are exactly anti-correlated. A value near zero implies no 
correlation. A CCF is straightforward to compute when the two time series of interest 
are uniformly sampled with the same time binning. However, within the context of 
astronomical time series, often the data are non-uniformly sampled owing to the various 
observing constraints associated with telescopes. For these unevenly sampled cases, the 
three most widely used forms of CCF are the interpolated cross-correlation function 
\citep[ICCF;][]{iccf}, discrete cross-correlation function \citep[DCF;][]{dcf}, and the 
z-transformed discrete cross-correlation function \citep[zDCF;][]{zdcf}. 

The ICCF is perhaps the most straightforward way to compute a CCF between two unevenly 
sampled light curves. In this method, for each specified value of time lag, one of the 
two light curves is shifted by that time lag and its flux values are interpolated 
(often linearly) at the times of the second light curve. Often the ICCF is 
computed twice, once by fixing the first light curve and shifting and interpolating the second light 
curve and vice versa \citep{whitepeter94}. The final ICCF is an average of the two CCFs  
and is thus not significantly biased towards the sampling of either of the two light 
curves. A DCF, on the other hand, only uses the actual data points without any 
interpolation \citep[see,][for more details]{dcf}. A zDCF is based on the DCF with several 
modifications. It has been argued that zDCF can tease out weaker cross-correlation signals 
from data than the traditional DCF \citep[see,][]{zdcf}. All three methods have been used
successfully within the context of AGN reverberation mapping but only rarely for TDE studies.

\begin{figure}[htp]
    \centering
    \includegraphics[width=\textwidth]{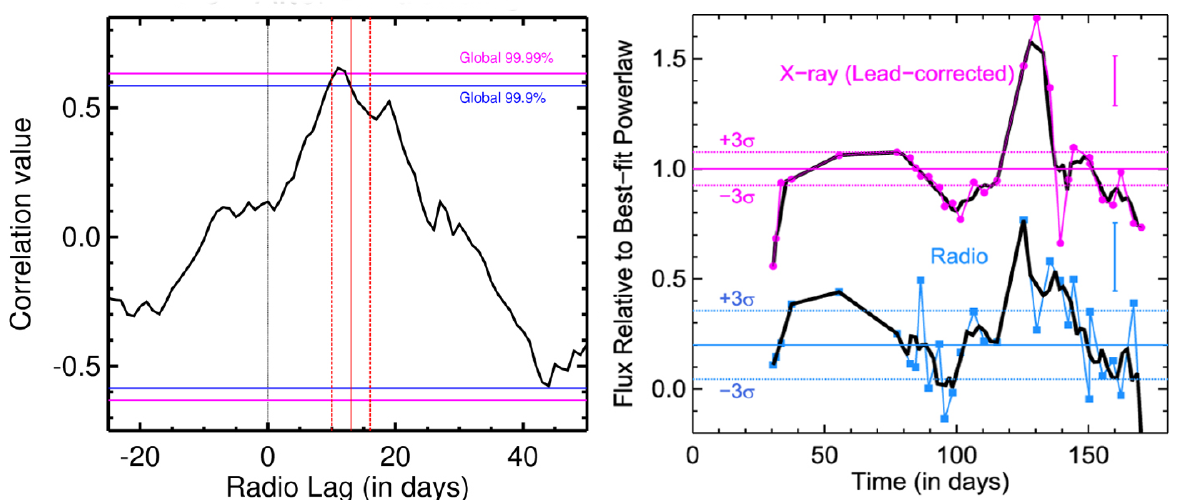}
    \caption{ {\it Left:} Cross-correlation function (CCF) between the soft X-ray (0.3-1 keV) and 16 GHz radio data from ASASSN-14li. The horizontal blue and magenta lines are the global 99.9\% and the 99.99\% white noise statistical confidence contours for a blind search between -25 and 50 days \citep[see Sec. 3.2 of ][ for more details]{pashvan14li}. The solid and the dashed vertical lines are the median and the 1$\sigma$ deviations (${13}_{-3}^{+3}$ days) derived by considering both the sampling and the measurement uncertainties in both the X-ray and radio light curves. {\it Right:} The relative X-ray (magenta) and radio (blue) light curves of ASASSN-14li obtained by dividing their corresponding best-fit power-law models. Both the light curves show the same variability pattern. The X-ray and radio fractional variability amplitudes on top of the power-law decay are 10\%$\pm$1\% and 16\%$\pm$1\%, respectively. The mean flux levels are shown by the solid horizontal lines while the $\pm$3$\sigma$ variability contours derived using the methodology described in \citet{rmero11} are indicated by dashed horizontal lines. The radio data have been vertically offset by -0.8. The solid black curves are running averages over a 10 day window (except when the gap between observations in longer than 10 days). Typical 1$\sigma$ measurement uncertainties are shown as vertical bars. The X-ray data points shown here have been interpolated to match with the radio epochs \citep[see Sec. 3.1 of ][ for more analysis details]{pashvan14li}.  (adapted from \citealt{pashvan14li}).
      }
    \label{fig:14liccf}
\end{figure}

\subsection{Radio and X-ray cross correlations}\label{sec:radiox_cc}
Because of its panchromatic brightness, ASASSN-14li was monitored in the radio and the 
X-ray (0.3-10 keV) wavebands roughly once every 3-10 days with a temporal baseline of a few hundred 
days \citep{pashvan14li}. ASASSN-14li's X-ray spectrum was soft with very weak emission above $\sim$1 keV. These data sets provided the first opportunity to search for reverberation signals on timescale of tens of days from a TDE. \citet{pashvan14li} extracted an ICCF between the X-ray and the 16 GHz radio emission from ASASSN-14li and found that the emission at 16 GHz lagged behind the soft/thermal X-rays by 
roughly 13 days (see Figure \ref{fig:14liccf}). Interestingly, both the fluctuations on top of the overall long-term 
decay and the long-term decay curves lagged by the same timescale. This suggested that 
the entire radio flux at 16 GHz was responding to variations in the soft X-ray (0.3-1.0 keV) 
brightness changes.

Based on this coupling between the radio and the soft X-rays and time-resolved radio 
spectral modeling, \citet{pashvan14li} suggested that the radio emission from 
ASASSN-14li originates from a sub-relativistic jet that is regulated by the inner 
accretion disk which emits X-rays. Moreover, they showed using the correlation 
between the two wavebands and independently via modeling the time-resolved radio 
spectral energy distributions, that the accretion rate and the jet power are 
linearly coupled, i.e., the newly formed accretion disk is efficiently regulating the 
radio jet \citep[for more details see Sec. 4.4 of][]{pashvan14li}. Such a strong coupling between the soft X-rays, i.e., inner accretion flow, and the radio emission has never been seen before in an AGN or a supernova and thus may be a means to distinguish TDEs from TDE imposters. However in order to establish this as a common TDE phenomenon, similar studies of several events in the future are necessary. Nevertheless, this work demonstrated the use of TDE radio and X-ray observations to probe the earliest phases of jet formation which is currently only poorly constrained by observations. 

\begin{figure}[t]
    \centering
    \includegraphics[width=\textwidth]{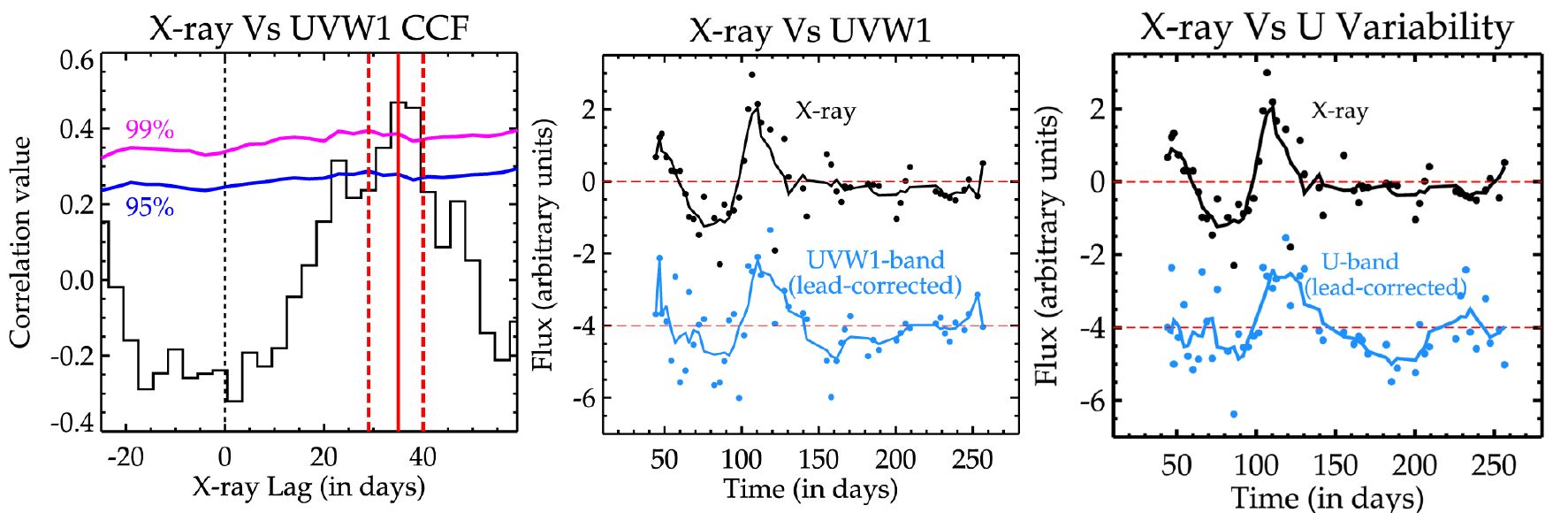}
    \caption{ {\it Left:} Cross-correlation function (CCF) between the soft X-ray (0.3-1 keV) and the UV (UVW1 filter of {\it Neil Gehrels Swift} observatory) light curves from ASASSN-14li. The blue and magenta lines are the statistical significance contours \citep[see ][ for more details]{Pasham:2017a}. The vertical red lines have the same meaning as in the left panel of Figure \ref{fig:14liccf}. {\it Middle and Right:} UV and the optical (U band) variability features are compared to the X-ray variations. The UVW1/U-band light curves are offset by 35 days and the X-ray light curve is interpolated at the lead-corrected UVW1/U epochs. In both panels the observed X-ray and UVW1/U light curves were de-trended by subtracting a smooth function, leaving only the variability features. The solid curves are a running average of five neighboring points, to guide the eye (adapted from \citealt{Pasham:2017a}).}
    \label{fig:14liuvccf}
\end{figure}

\subsection{Optical/UV and X-ray cross correlations }\label{sec:uvx_cc}
\begin{figure}[t]
    \centering
    \includegraphics[width=\textwidth]{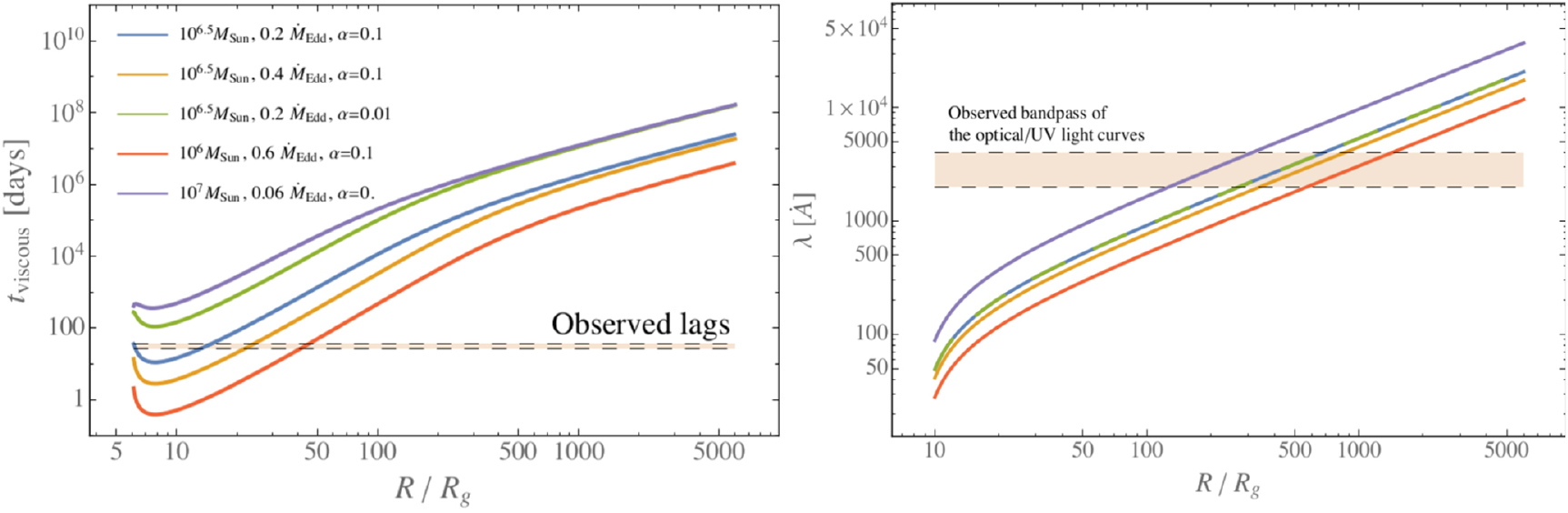}
    \caption{ The dependence of the local viscous timescale (left) and the wavelength of peak emission (right) on the radial distance from the black hole in a Shakura-Sunyaev accretion disk for a set of black hole masses, Eddington ratios and $\alpha$-viscosity parameter values. The observed timescale of $\sim$30 days is inconsistent with a circular disk and argues for an accretion flow in ASASSN-14li that is distinct from AGN accretion disks (adapted from \citealt{Pasham:2017a}).}
    \label{fig:14lisscomp}
\end{figure}

\begin{figure}[t]
    \centering
    \includegraphics[width=0.5\textwidth]{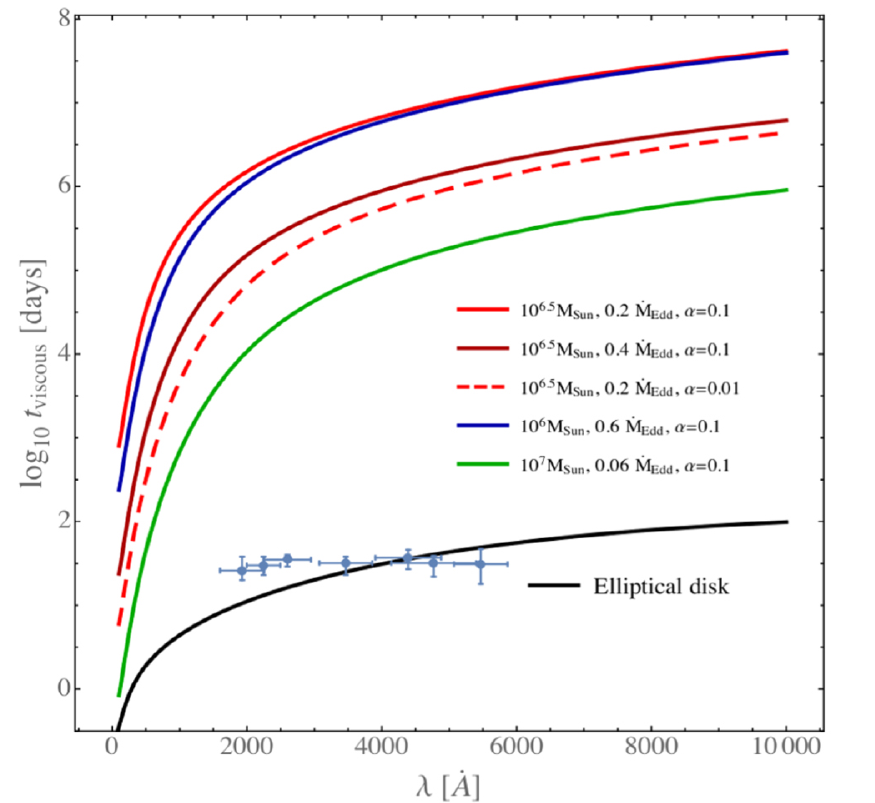}
    \caption{ The colored curves (green, blue and red) show the dependence of viscous timescale on wavelength of emission for a standard, circular thin disk. The observed time lags, shown as blue data points, are orders of magnitude faster than expected from a circular thin disk. The viscous time lags in a simplistic elliptical accretion disk described in \citet{g14} (black curve) are shorter and are broadly consistent with the observed optical/UV--X-ray time lags in ASASSN-14li (adapted from \citealt{Pasham:2017a}).}
    \label{fig:14liellip}
\end{figure}

In addition to radio--X-ray correlation studies, the inclusion of the optical and the UV wavebands in TDE photometric reverberation analysis can reveal insights into the physical origin of optical/UV light. Within the context of AGN, time lags between the X-ray and the optical/UV variations can arise from the following various physical scenarios \citep[e.g.,][]{mchardy:lagdesc}. Firstly, 
X-rays from near the black hole can scatter off material in the outer accretion disk or 
any other surrounding medium, lose energy, and get reprocessed into lower-energy 
optical/UV photons \citep{edelson:2015}. In this case, fluctuations in the X-rays would 
lead the optical changes by a few days to a few tens of days depending on the distance 
(and thus the light travel time) to the outer disk/reprocessing region 
\citep{morgan:disksize}. Alternatively, if the optical and the UV emission is dominated 
by the thermal emission from the accretion disk, two kinds of correlations/lags are 
possible. (1) If the X-rays are produced in a Corona by Compton up-scattering of the 
thermal near-UV ``seed'' photons \citep{reynoldsnowak:2003}, then the changes in the X-ray flux 
would lag behind the near-UV fluctuations by a fraction of a day to a few days depending 
on the size of the corona and the inner accretion disk \citep[which are determined by the 
black hole mass, e.g., ][ and accretion rate]{arevalo:2005}. (2) Also in this scenario, 
because the optical/UV originates directly from the disk, accretion rate perturbations 
can manifest as variations in optical and UV emission. As these perturbations propagate inwards in the disk on the local viscous timescale \citep{arevalo:2008}, they 
first trigger changes in the optical band followed by the UV and finally X-rays. Thus, we may expect the optical/UV emission to lead the X-rays by a few tens to millions of days depending on the black hole mass and the accretion rate \citep{ss73}, see Figs. \ref{fig:14lisscomp} and \ref{fig:14liellip}. 

Due to its early identification as a compelling TDE, ASASSN-14li was monitored at a cadence of once every 3-7 days in the optical/UV bands for about 250 days. Motivated by optical/UV--X-ray photometric reverberation mapping of AGN \citep{edelson:2015, mchardyngc4593:2018}, \citet{Pasham:2017a} performed cross-correlation analysis between the soft X-ray and the several of ASASSN-14li's available optical/UV light curves. {They found evidence for the presence of similar flux variations in both the X-ray and optical/UV bands.} In addition, the variability pattern in X-rays lagged behind those in the optical and the UV bands by roughly 32 days (see Figure~\ref{fig:14liuvccf}).

Based on the direction of the time lag (X-rays lagging the optical/UV), it appears that X-ray reprocessing is not the dominant source of optical/UV emission in ASASSN-14li. Furthermore, the seed photon scenario described above is also unlikely because the expected UV lead times for a 10$^{6.5\pm0.6}$ M$_{\odot}$ black hole--implied from host galaxy properties--are only a few thousand to a few tens of thousand seconds, which is an order of magnitude shorter than the observed lead time. Also, in ASASSN-14li, there is no evidence for a Corona (non-thermal emission) in the high quality X-ray spectra obtained during the first $\sim$300 d of monitoring. 

The observed lags are also unlikely to be due to viscous time delays in a ``standard'' circular Shakura-Sunyaev accretion disk \citep{ss73}. ASASSN-14li's peak bolometric luminosity of $\sim 10^{44}$ erg\,s$^{-1}$ \citep{Holoien:2016b} implies a peak Eddington ratio of 0.2$^{+0.7}_{-0.2}$~ $\dot{M}_{\rm Edd}$ (where $\dot{M}_{\rm Edd}$ is the Eddington rate) for a black hole in the mass range of ${10}^{6.5\pm 0.6}$ ${M}_{\odot}$ \citep[e.g.,][]{Holoien:2016b,pasham:14liqpo}. At such sub-Eddington rates, the newly formed accretion disk can be described by a geometrically thin, optically thick disk model of \citet{ss73}. For the observed lag timescale of $\sim$30 days to correspond to viscous delay in a disk the optical/UV emission should originate from within a radius of 50 $R_{g}$ (see the left panel of Figure \ref{fig:14lisscomp}). However, the bulk of the optical/UV emission in the observed {\it Swift} bandpass (1500-6000 \AA) originates from beyond a few hundred gravitational radii (see the right panel of Figure \ref{fig:14lisscomp}) and at these larger radii the viscous timescale and thus the time lags are orders of magnitude longer. This discrepancy between the emission radius, emission bandpass and corresponding viscous timescales rules out the standard, thin, circular disk solution.


This discrepancy can be resolved in a few of ways. First, the disrupting black hole could be a few orders of magnitude smaller than 10$^{6}$ $M_{\odot}$, i.e., a so-called intermediate-mass black hole. This would make the accretion flow super-Eddington and produce slim disks \citep{slimdisk} which can yield faster viscous timescales, i.e., comparable to the optical/UV--X-rays lags observed in ASASSN-14li. Alternately, an elliptical accretion disk would also shorten the viscous timescales (see Figure \ref{fig:14liellip}). 

In any case, it is clear that at least in the quintessential TDE ASASSN-14li the optical/UV emission vary with respect to the central X-rays in a unique manner. It remains to be seen if this behavior is typical in TDEs.  If so, these distinguishing ``echo'' signals could be a means to separate genuine TDEs from imposters, viz., AGN flares and unusual supernovae. {We note that cross-correlation studies like the ones carried out for ASASSN-14li require high-cadence monitoring over a long temporal baseline (once every few days for several months) in multiple wavebands, and therefore may not be feasible in many cases.}


\begin{table}[b]
\caption{Summary of the currently known TDE quasi-periodic oscillations (QPOs).}
  \begin{threeparttable}
     \begin{tabular}{P{2.5cm} P{2.15cm} P{1.25cm} P{4.2cm}}
        \hline\\[-7pt]
        Flare name & Centroid Frequency$^{\dagger}$ & Fractional rms$^{*}$ & Notes \\
        \hline\hline\\[-7pt]
        SwJ1644+57 & 4.8$\pm$0.3 mHz & 3-4\% &  QPO was present only within the first two weeks after the flare was discovered.\\ 
  & &  &\\ 
  \hline \\[-7pt]
 2XMM J123103.2+110648 & $\sim$7.3$\times$10$^{-5}$ Hz & 20-60\% & The optical spectrum is unusual compared to typical TDEs. Could be an AGN flare.\\ 
  & &  &\\
  \hline \\[-7pt]
 ASASSN-14li & 7.65$\pm$0.4 mHz & 3-40\% & Stable QPO lasting for at least 500 d after the flare was discovered.\\ 
        \hline \\[-7pt]
     \end{tabular}
    \begin{tablenotes}
      \small
      \item $^{\dagger}$The centroid frequency of the X-ray QPO. 
      \item $^{*}$The fraction root-mean-squared amplitude of the QPO in percentage of mean flux.
    \end{tablenotes}
  \end{threeparttable}
  \label{table:qpos}
\end{table}

\section{X-ray Quasi-Periodic Oscillations from TDEs}\label{sec:qpo}

Quasi-periodic oscillations (QPOs) in the X-ray flux with centroid frequencies between a few mHz to several hundreds of Hz have now been observed from several stellar-mass black hole binaries \citep{mcclinronbook}. These are systems where the central compact object has a mass between 5 and 20 M$_{\odot}$ and draws material from a non-degenerate companion star. X-ray QPOs in stellar-mass black hole binaries often appear when the black hole goes into an outburst during which the overall X-ray flux can increase by orders of magnitude for a brief duration of a few weeks to months \citep[e.g.,][]{homan:2005, ron:2002}. While the exact origin of these X-ray modulations is unclear, they are generally thought to be associated with the dynamics of hot gas in the innermost regions of the accretion flow (see \citealt{mcclinronbook} for a review). In some systems, the spins of the central black holes have been estimated under the assumption that the QPOs are associated with test particle orbits predicted from general relativity \citep{franchini:2017, motta:2014}.

\begin{figure}[tb]
    \centering
    \includegraphics[width=\textwidth]{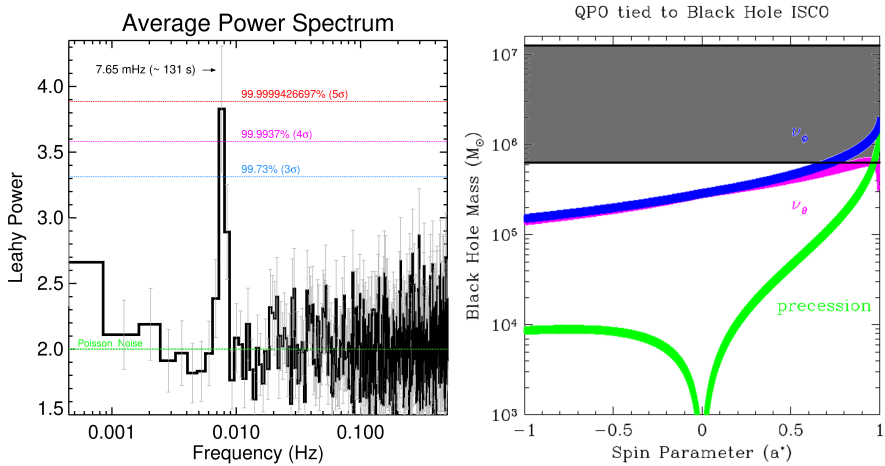}
        \caption{{\it Left:} Average X-ray (0.3-1 keV) power density spectrum of ASASSN-14li using data segments spread over 500\,d (adapted from \citealt{pasham:14liqpo}). The global statistical confidence levels and the Poisson noise level are shown as dashed horizontal lines. The strongest features in the spectrum is at 7.65$\pm$0.4 mHz. {\it Right:} Black hole mass vs dimensionless spin parameter contours estimates by assuming the QPO is associated with test particle orbits at the innermost stable circular orbit (ISCO). Contours corresponding to three particle frequencies: Keplerian frequency ($\nu_{\phi}$, blue), vertical epicyclic frequency ($\nu_{\theta}$, magenta) and Lense-Thirring precession ($\nu_{\phi}$ - $\nu_{\theta}$, green) at the ISCO are shown. At the ISCO the radial epicyclic frequency ($\nu_{r}$) is zero and the periastron precession frequency ($\nu_{\phi}$ - $\nu_{r}$) is thus equal to the Keplerian frequency. The widths of these contours reflect the  QPO's width of 0.7~mHz (upper limit). The shaded grey rectangular area represents ASASSN-14li's black hole mass range (10$^{5.8-7.1}$~$M_{\odot}$) estimated from its host galaxy scaling relations. Within this mass range, the only formal solutions are the ones that require the spin parameter to be greater than 0.7. Choosing a radius larger than the ISCO only pushes this lower limit to a higher value.}
    \label{fig:14li}
\end{figure}

A wealth of X-ray timing studies now suggest that supermassive black holes (SMBHs) are scaled-up versions of stellar-mass black holes in some aspects \citep[e.g.,][]{mchardy:bhunification, markowitz:agnpds}. For example, it has been shown that the X-ray power density spectra of both SMBHs in AGN and stellar-mass black holes are qualitatively similar and that certain timescales within the power spectra scale inversely with the black hole mass after correcting for the accretion rate (\citealt{mchardy:bhunification, markowitz:agnpds}; but also see \citealt{donemark:2005}). Under this black hole unification paradigm, even SMBHs should exhibit QPOs but with frequencies that are 10$^{5-7}$ times lower depending on the exact SMBH mass. However, in spite of several searches, robust evidence for X-ray QPOs from SMBHs in AGN has been scarce (see, however, \citealt{gierlinski:qpo, alston:agnqpo}). This could be because of a dearth of highly sensitive observations or that QPOs are intrinsically absent in steadily accreting SMBHs (see \citealt{vaughan:whereareagnqpos} for details). 

On the other hand, TDEs are different from steadily accreting SMBHs and to some degree are similar to stellar-mass black hole binaries when they go into outburst. This is because in both stellar-mass black hole outbursts and TDEs starving black holes are suddenly fed at a high rate. Thus, the natural expectation is that some of the same phenomena, viz., X-ray QPOs, that occur in stellar-mass black hole outbursts should also appear in TDEs. There have been three cases where X-ray QPOs have been reported from TDE (candidates). These include the relativistic TDE SwJ1644+57 \citep{reis:tdeqpo}, the super soft X-ray flare 2XMM J123103.2+110648 \citep{Lin:2015a}, and the thermal TDE ASASSN-14li \citep{pasham:14liqpo}. The observed properties of these three QPOs are summarized in Table \ref{table:qpos}.


\begin{figure}[tb]
    \centering
    \includegraphics[width=\textwidth]{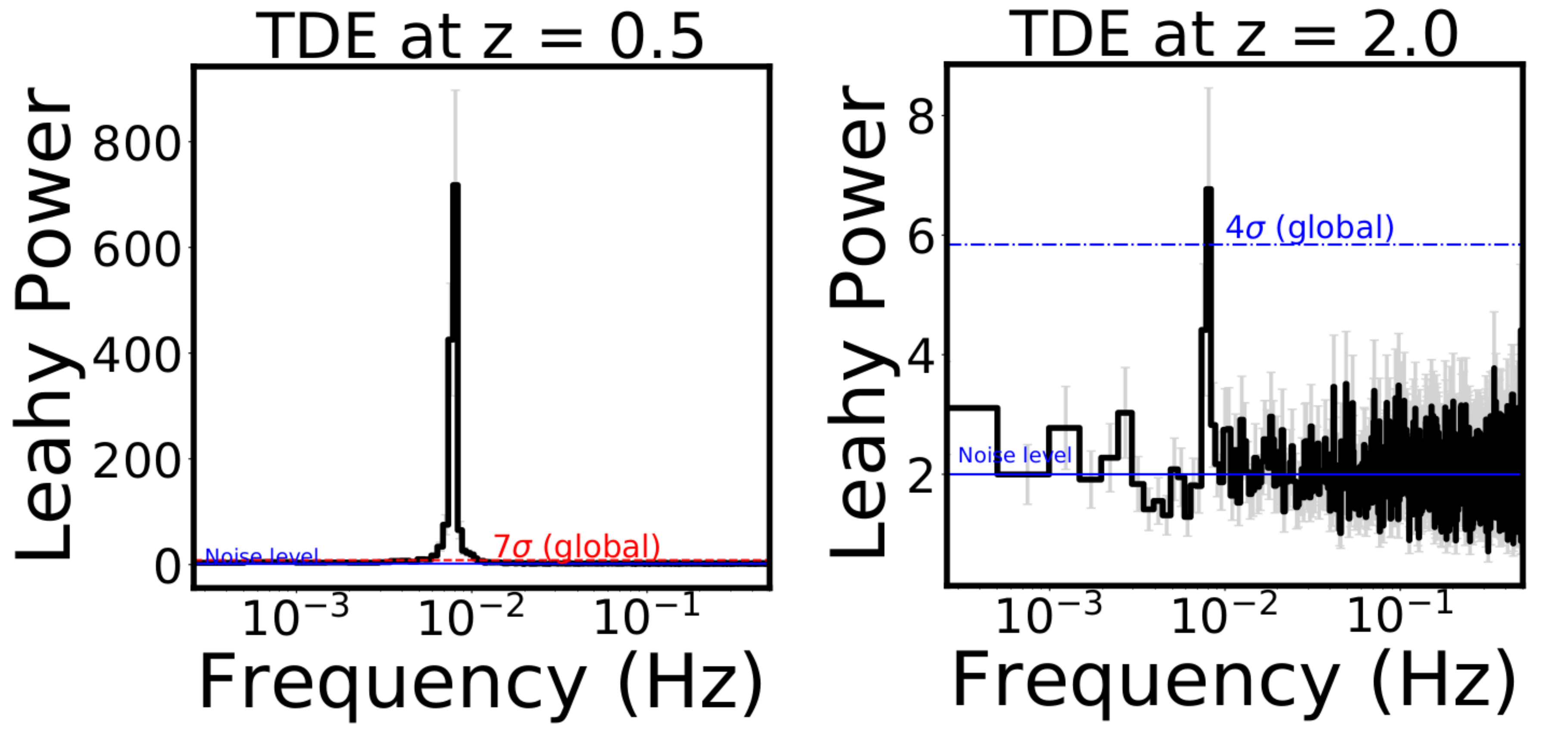}
    \caption{Simulated power density spectra of an ASASSN-14li-like TDE at various redshifts using Athena's Wide Field Imager (WFI). For each of these simulations we used an exposure time of 32 ks and a QPO fractional rms amplitude of 40\% (similar to ASASSN-14li) whose value remains constant with energy within 0.3-2 keV bandpass. Statistical confidence contours and the Poisson noise level of 2 are shown as dashed horizontal lines. The time dilation effect due to the redshift and the potential well of the source on the centroid frequency of the QPO was neglected as it will not alter the results significantly. Assuming ASASSN-14li-like behavior, TDE X-ray QPOs could be detected out to a redshift of roughly 2 with a meter-class X-ray telescope.}
    \label{fig:athenasims}
\end{figure}

 The case of ASASSN-14li is particularly interesting. Because of its widespread interest ASASSN-14li was monitored for several years at multiple wavelengths. In particular, the flare was detectable with {\it Swift}’s X-Ray Telescope (XRT) for multiple years after its discovery and several deeper exposures were obtained with {\it Chandra} and {\it XMM-Newton} during this time. An average power density spectrum of the {\it Chandra} and the {\it XMM-Newton} data segments spread over the first 500 d reveals a prominent feature at a frequency of 7.65$\pm$0.4 mHz (Figure \ref{fig:14li}, left panel). The feature was surprisingly stable for over 300,000 cycles and had a fractional root-mean-squared amplitude of 40\% around day 450. The stability for such large number of cycles while the flux changed by an order of magnitude suggests an origin that is tied to the fundamental properties, viz., the mass and the spin of the disrupting black hole \citep{pasham:14liqpo}. Using a black hole mass implied from host galaxy properties, i.e., based on the scaling laws between the black hole mass and the host’s stellar velocity dispersion and the luminosity, and associating the QPO period to the various particle frequencies predicted from general relativity, \cite{pasham:14liqpo} constrained the dimensionless spin parameter of the disrupting black hole to be at least 0.7 (Figure \ref{fig:14li}, right panel). This study presented a new means to constrain spins of several SMBHs in the future when they tidally disrupt stars. On the other hand, the contours in the right panel of Figure \ref{fig:14li} can also be used to set an absolute upper limit on black hole mass. For ASASSN-14li this upper limit corresponds to 2$\times$10$^{6}$ $M_{\odot}$. For future events with QPOs this could be a straightforward technique to identify the long-sought intermediate-mass black holes (masses in 10$^{2-5}$ $M_{\odot}$) whose QPO timescales would be less than a few tens of seconds.

Historically, not many X-ray bright TDEs have been followed-up with highly sensitive X-ray observations  
as a majority of them were discovered in archival searches
(e.g., \citealt{komossabade:1999, esquej:xmmtdes} and see \citealt{komossa:tderev} and references therein). 
But, several of the events that did have detailed X-ray observations showed QPOs. While the number of such X-ray TDEs with QPOs is currently small, it remains plausible that QPOs may have been missed in several earlier TDEs. If ASASSN-14li is indeed a ``poster child'' for TDEs and the observed X-ray QPOs are typical in these systems, 
then this opens a new opportunity to use TDEs to build a census of black hole mass. An X-ray observatory with an effective area of the order of a square meter will be transformational for this purpose. For instance, simulations using spectral responses of the Wide Field Image (WFI) and the X-ray Integral Field Unit (X-IFU) on board the anticipated meter-class X-ray observatory Athena suggest that ASASSN-14li like QPOs could be detected out to a redshift of 2 (Figure \ref{fig:athenasims}). 

The prospect of detecting QPOs from TDEs at cosmological distances is very exciting because it would allow us to use these signals to constrain spins of a large number of SMBHs and also construct SMBH spin distributions at various redshifts. On the other hand, the QPOs themselves are excellent probes of accretion at these redshifts. Spin distributions are especially important to constrain the models for growth of SMBHs. For example, it has been predicted that if SMBHs grew primarily via mergers with other black holes, their spin distribution should be skewed towards low spin \citep[e.g.,][]{bertivolon:smbhspin}. This is because mergers tend to deposit angular momentum along random orientations which tends to generally spin down the remnant hole. On the other hand, growth by steady accretion tends to add angular momentum along the spin axis of the hole and thus tends to spin up the SMBH. Thus TDE QPOs and the resulting spin constraints could provide a novel opportunity to understand the dominant growth channel of SMBHs.


\section{X-ray spectroscopy: relativistic reverberation}\label{sec:xreverberation}

X-ray light echoes probe the smallest scales around the black hole. X-ray reverberation is commonly seen in AGN systems, most of which are sub-Eddington systems consisting of an isotropically-emitting X-ray corona that irradiates the thin accretion disc (e.g. \citealt{demarco13, uttley14}). This produces fluorescence lines with a time delay corresponding to a light travel distance of a few gravitational radii (e.g. \citealt{zoghbi12, kara16a}). The reverberation lags found in AGN can probe the extent of the corona and the inner radius of the thin accretion disc (e.g. \citealt{wilkins16,taylor18}).

Thus far, only one TDE is bright and variable enough (on timescales of hundreds to thousands of seconds) to enable the detection of X-ray time lags. This source is Swift J1644+57, the most well-observed hard X-ray TDE. In this source, \citet{kara16b} discovered an emission line at 8~keV, which was seen to lag behind the continuum by 100~seconds. If associated with {the Fe~K$\alpha$ line} (the line with the highest fluorescent yield in the X-ray spectrum), then the line must be highly blueshifted.

The extremely high isotropic luminosity and the detection of strong radio emission from Swift J1644+57 suggest that this is a TDE with a jet along our line of sight, and a face-on accretion flow. The highly blueshifted iron K line and corresponding reverberation lag can be well described by reprocessing off of a Compton thick, outflowing disc wind with a velocity of $>0.1c$ and $\sim 30 \deg$ opening angle \citep{kara16b, lu17, thomsen19}. The iron~K lag is marginally better  described ($>98\%$ confidence) by an asymmetrically broadened line  with some gravitational redshifted over a symmetric Gaussian. This suggested that the reprocessing occurred close to the black hole, and therefore at least some of the X-ray emitting continuum is isotropically emitting \citep{kara16b}. This was also supported by the short 100~s lag, { though, \citet{lu17} point out that if the iron line band also contains emission from the primary continuum source (an effect known as `dilution'), then} the iron~K reprocessing region could be further from the black hole, and thus do not rule out that the X-rays are produced in the slow sheath of a relativistic jet.  

To conclude, we have shown that X-ray reverberation lags probe the accretion flow structure around black holes, and can be used to probe the velocity and geometrical structure of ultrafast outflows in TDEs. 

\section{Concluding remarks and outlook}\label{sec:discussion}
Since the strength of a reverberation signal depends on the availability of material to reflect or reprocess the transient emission, we anticipate these signal are not universal in TDEs. Indeed, dust echoes are not detected for every TDE \citep[e.g.][]{van-Velzen:2016a,Jiang21} and the extreme coronal lines seen in SDSS\,J0952+2143 (Figure~\ref{fig:J0952-ecle-broadband}) {and several other SDSS sources} have never been observed in follow-up optical spectra of TDEs detected in other imaging surveys. Furthermore, since not every TDE is X-ray bright \citep[e.g.][]{Auchettl:2017a}, the X-ray cross-correlation and timing signatures discussed this review also remain relatively rare, with detections in just one (either ASASSN-14li or Swift~J1644) or a handful of sources: see Table~\ref{table:qpos}.  

While they are not common to the full TDE population, the detection of reverberation signals is important since they provide a novel way to study the TDE emission mechanism and these signals can be used to select TDEs that could be missed by traditional photometric searches for TDEs. The events selected based on the detection of galaxies with extreme coronal lines are well-suited to recover a subset of TDEs that occur in gas rich circumnuclear environments, while a search for IR echoes can reveal TDEs in galaxies with a large amount of dust extinction to the center \citep[e.g.,][]{Wang18,Mattila18}. We should note that IR follow-up observations of TDE candidates in AGN have yielded dust echoes that are an order of magnitude more luminous compared to IR echoes in TDEs from quiescent galaxies (Table~\ref{tab:ir}), implying that a blind search for IR echoes {could be} dominated by flares in AGN (see also the discussion at the end of \ref{sec:IRcompare}). 

Both the IR echoes and the detection of coronal lines can be used to probe the energy produced at EUV wavelengths (which cannot be observed directly due to absorption by neutral hydrogen). In most cases, the observed optical/UV or X-ray luminosity is lower than the energy inferred from the reverberation signals. This provides support for models that explain the optical/UV emission of TDEs with an accretion disk \citep{Cannizzo90,Dai18,vanVelzen18_FUV,Mummery20}, since for black holes in the mass range $M_{\rm BH}\sim 10^{6-7}\,M_\odot$ these disk are expected to emit most of their energy in the EUV regime, see also the \emischap of this volume. 

While the IR echoes are agnostic to the source that is heating the dust, the cross-correlation signals detected for ASASSN-14li (Section~\ref{sec:crosscorr}) appear to be unique to TDEs and can thus be used to help discriminate TDEs from impostors such as AGN flares, see the \impostchap of this volume for further discussion. In particular the detection of a QPO in the X-ray light curve of ASASSN-14li is truly spectacular, providing new ways to constrain the spin of the black hole that disrupted the star. The relativistic reverberation signal (Section~\ref{sec:xreverberation}) detected for the jetted TDE Swift~J1644+57 is another unique signature which probes the geometry of the inter accretion flow and the base of the jet. 

Clearly the detection of more multi-wavelength cross-correlations and QPOs is very much anticipated. The current detection rate of such events is $\sim 1$ per decade. While recently the detection rate of optical TDEs has increased significantly, to dozens of new sources per year \citep[e.g.,][]{vanVelzen20}, few of these have well-sampled X-ray light curves similar to ASASSN-14li (one notable exception is AT2018fyk,see \citealt{Wevers19a,Wevers21}). This lack of useful X-ray light curves is mainly explained by the low X-ray flux of optically-selected TDEs, but is also due to a scarcity of X-ray follow-up resources. An important improvement would be a new X-ray mission that is dedicated to high-cadence observations, such as Einstein Probe \citep{Yuan15}. As discussed in section~\ref{sec:qpo}, X-ray observations with a meter-class X-ray observatory such as Athena would be able to detect QPOs from TDEs to $z\sim 2$, thus constraining the spin of black holes in the cosmic epoch when they are expected to be rapidly growing.

At the time of writing, the WISE satellite continues to monitor the near-IR sky (3-4 $\mu$m) with a cadence of 6 months, thus providing an excellent dataset that can be used to obtain detections (or upper limits) of dust echoes for every transient detected in the last decade. The (potential) detection of dust reprocessing using ground-based $i$-band observations \citep{Stein20} {and $K$-band observation \citep{Mattila18,Kool20}} is also an encouraging result, because this allows for a much higher cadence than the scanning pattern of WISE. In the near future, observations with JWST could yield a detailed spectroscopic view of dust heated by TDEs. 




\bibliographystyle{aps-nameyear}      

          
\bibliography{general,extra}

\end{document}